\documentclass[dvips]{acta}
\usepackage{supertabular,lscape,epsfig}
\usepackage{amssymb}
\usepackage{amsmath}

\SetPages{1}{1}

\SetVol{59}{2009}

\begin{document}

\begin{Titlepage}
\Title{Double-Overtone Cepheids in the Large Magellanic Cloud}

\

\Author{W.A.~~D~z~i~e~m~b~o~w~s~k~i$^{1,2}$~~and~~R.~~S~m~o~l~e~c$^2$}
{$^1$Warsaw University Observatory, Al.~Ujazdowskie~4,
00-478~Warsaw, Poland\\
e-mail: wd@astrouw.edu.pl\\
$^2$Copernicus Astronomical Center, ul.~Bartycka~18, 00-716~Warsaw, Poland}

\Received{Month Day, Year}
\end{Titlepage}

\Abstract{One of the most interesting results from the OGLE-III study of
the LMC Cepheids is the large number of objects that pulsate
simultaneously in the first and second overtone (denoted 1O/2O).
Double-mode Cepheids yield important constraint on stellar evolution
models. We show that great majority of the LMC 1O/2O Cepheids  have
masses $M=3.0\pm 0.5M_\odot$. According to current stellar evolution
calculations, these masses are lower than needed for the blue loop
in the helium burning phase to reach the instability strip. On the
other hand, we found most of these stars significantly overluminous
if they are crossing the instability before helium ignition.
A possible solution of this discrepancy is to allow for a large
overshooting from the convective core in the main sequence phase. We
also discuss origin of double-mode pulsation. At the short period
range we find two types of resonances that are conducive to this
form of pulsation. However, at longer periods, it has a different
(non-resonant) origin.}{}

\section{Introduction}

Double-mode (called also Beat) Cepheids are important objects for
stellar physics. The diagnostic tool they provide are Petersen
diagrams, where shorter to longer period ratio is plotted against
log of the longer period. For nearly twenty years after they were
introduced by Petersen (1973), the discrepancy between observational
and calculated period ratio constituted a great challenge to stellar
evolution theory. The major revision of stellar opacity calculation
in the late 1980ties has been inspired by Simon's (1982) suggestion
based mainly on this discrepancy.

Only a small fraction of Cepheids develop double-mode pulsation.
According to Soszy\'nski et al. (2008b, hereafter S2008), it occurs only in
less than 8 percent of the LMC Cepheids. It is difficult to
develop this form of pulsation not only in nature but also in nonlinear
modeling. Koll\'ath et al. (1998) (see also Szab\'o, Koll\'ath and
Buchler 2004) managed to obtain models exhibiting sustained double-mode pulsation but, as Smolec and Moskalik (2008b) argued, their
treatment of turbulence was incorrect.

So far attention has been mostly focused on Cepheids with excited
fundamental (F) and first overtone (1O) modes. Readers are referred,
in particular, to recent papers by Buchler and Szab\'o (2007) and
Buchler (2008) on application of the F/1O Cepheid data as probes of
stellar and stellar system metallicity. Surprisingly, data on the more
frequent 1O/2O double-mode pulsators did not attracted much
attention. In the LMC there are more than three times as many 1O/2O than
F/1O Cepheids (S2008). The present paper is devoted almost solely to the
former type.

\section{OGLE-III data for the double- and triple-overtone Cepheids in the LMC}
\begin{figure}
\centering
\includegraphics[width=\textwidth,clip]{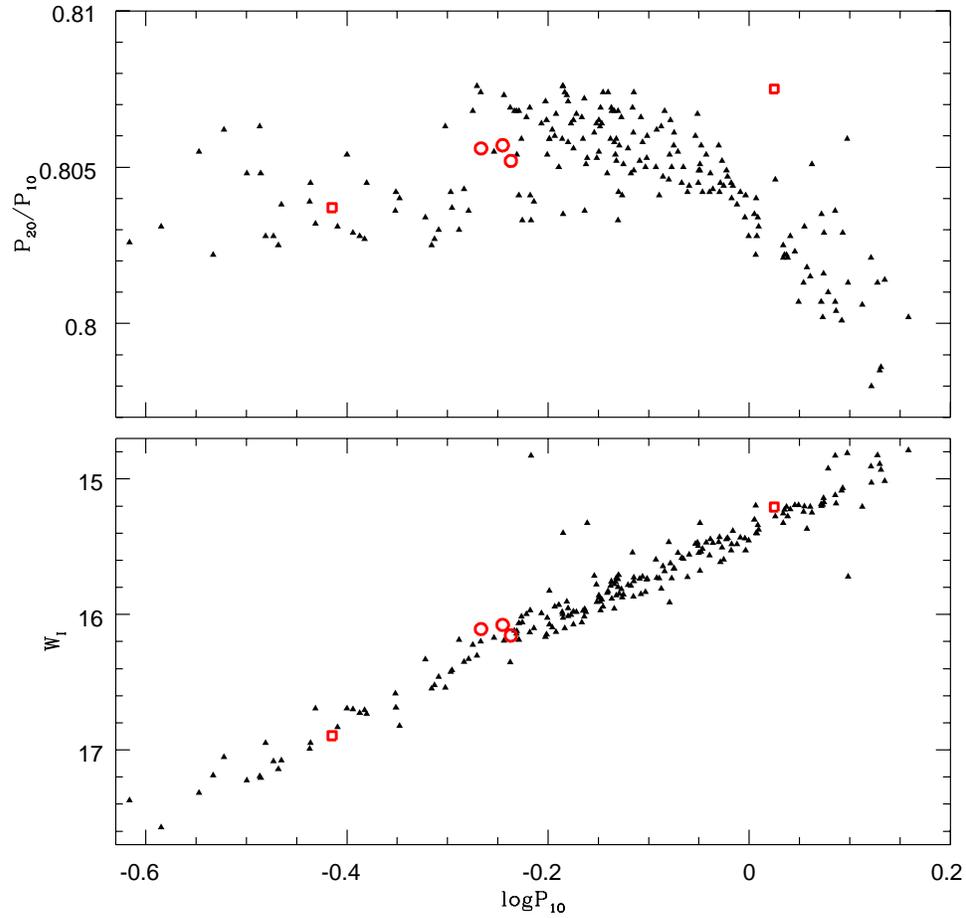}
\caption{ The $P_{2\rm O}/P_{1\rm O}$ period ratio (top) and Wesenheit index,
$W_I=I-1.55(V-I)$, (bottom) as function of the first overtone
period (days) for 203 1O/2O (solid triangles), two F/1O/2O (open
squares), and three 1O/2O/3O (open circles) Cepheids in LMC.}
\end{figure}
\begin{figure}
\centering
\includegraphics[width=\textwidth,clip]{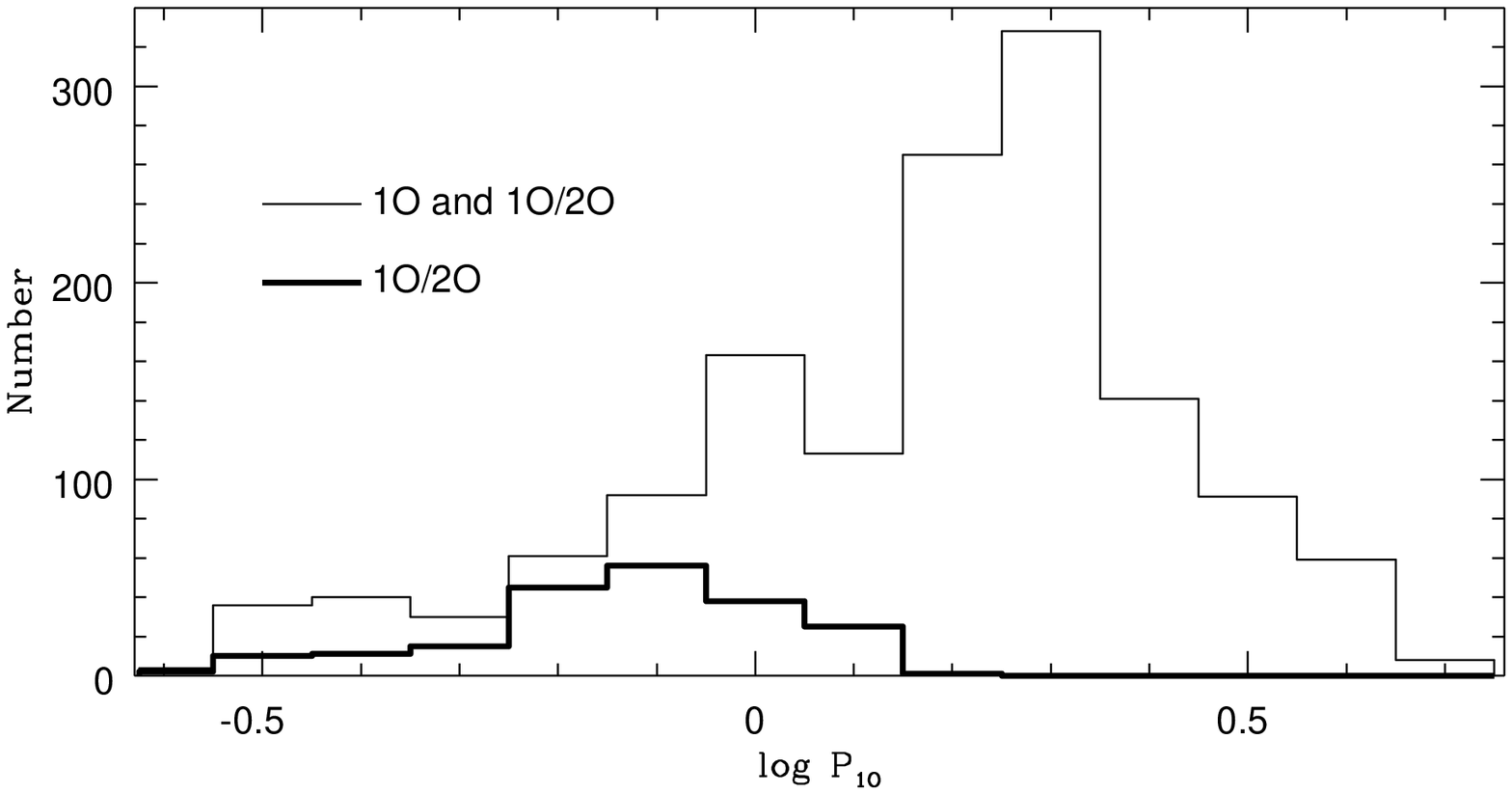}
\caption{ Period distribution of the 1O/2O and 1O Cepheids in the LMC according to S2008 data.
Total numbers of the 1O/2O and 1O objects are 203 and 1228, respectively}
\end{figure}
The Petersen diagram for the 203 LMC 1O/2O Cepheids is shown in
the upper panel of Fig.\,1. This is an expanded section of Fig.\,2 in S2008,
where also the $P_{1\rm O}/P_F - \log P_F$ relation was shown. The
patterns of these two relations differ. The latter shows a simple
monotonic decrease. In our Fig.\,1 we see a barely marked increasing
tendency up to $\log P_{1\rm O}\approx-0.1$, followed by a sharp
decline. The lower panel shows the period-luminosity relation (PL)
for the same objects. The relation, in which the Wesenheit index is
used as a reddening-free measure of luminosity, is part of the
general PL relation for the first overtone Cepheids in the LMC
depicted in the bottom panel of Fig.\,6 in S2008.

In each panel, we put also five points corresponding to
triple-overtone objects (Soszy\'nski et al. 2008a). The three at
$\log P_{1\rm O}\approx-0.25$ are 1O/2O/3O pulsators. The triple mode
nature of two of them was first established with OGLE-II data by
Moskalik et al. (2004). The three periods were then used by Moskalik
and Dziembowski (2005) to construct seismic models of the object.
The two F/1O/2O objects are widely separated in period.

Soszy\'nski et al. (2008a) found also two Cepheids with excited first and third overtone, showing no traces
of the second overtone. These 1O/3O pulsators have $\log P_{1\rm O}=-0.277$ and -0.244. The $W_I$ indices are
within the band of the PL relation for the 1O/2O Cepheids.

Nearly half (101) of the 1O/2O Cepheids occur in the $\log P_{1\rm
O}$ range [-0.25,-0.05], that is, the 0.56-0.89 d period range,
where it is the most common form of Cepheid pulsation in the LMC.
Comparing histograms in Fig.\,2, we see that there is much more of
1O Cepheids but not in this period range. There is no such a
preferred range for the F/1O double mode Cepheids. In OGLE-III data
for the LMC the number ratio of 1O/2O to F/1O Cepheids is 3.2. The
corresponding ratio from the EROS-2 data is only 1.8 (Marquette et
al. 2009). This large difference is likely due to lower amplitudes
of the 1O/2O Cepheids and the lower peak detection threshold in
OGLE-III data\footnote{We thank Igor Soszy\'nski for this
explanation.}.
\section{Masses, metallicity an evolutionary status of the 1O/2O Cepheids in LMC}
Distribution of Cepheid periods is governed by various factors,
such as recent history of star formation, time spend in the
instability strip (IS), $\tau_{\rm IS}$, and mode selection.
All these factors have been discussed by Alcock et al. (1999) in the context
of the MACHO data on the LMC Cepheids. The
value of $\tau_{\rm IS}$ strongly depends on star mass and
evolutionary status. For stars with intermediate masses,
the first crossing of the IS takes place in the post-main sequence phase
before helium ignites and it is by
two orders of magnitude faster than subsequent crossings, which
occur in the core helium burning phase (e.g. Alibert et al. 1999).
Naturally, most of Cepheids are expected to be in the latter phase
and have masses close to the minimum value, $M_{\rm mIS}$, for the blue loop
of the evolutionary track in the HR diagram to reach up to the IS.
\begin{figure}
\centering
\includegraphics[width=\textwidth,clip]{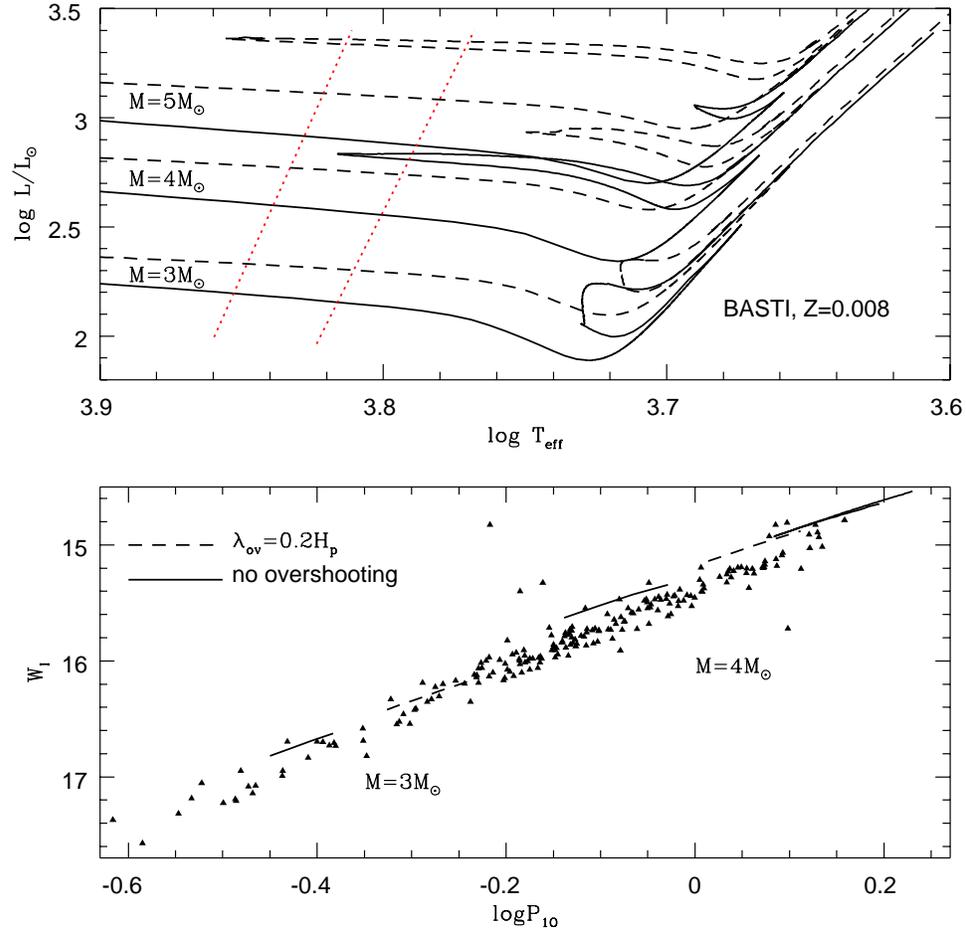}
\caption{The top panel shows evolutionary tracks from Pietriferini et al. (2004, BASTI files) calculated
neglecting (solid lines) and including (dashed lines) overshooting from convective core. The range
of simultaneous instability of the first and second overtones (red dotted lines) is from our calculations.
In the bottom panel the corresponding (only at $M=3 \mbox{ and }4M_\odot$) P-L relations  are compared with the OGLE-III data for the 1O/2O
Cepheids. The two segments corresponding to the second and third crossing of the IS at $M=4M_\odot$
nearly coincide in this panel.}
\end{figure}
In the OGLE-III sample containing over three thousands objects, we
expect to find a significant number of stars during fast crossing
of the IS. Furthermore,  there are problems with explaining short period
Cepheids as core helium burning objects (Alibert et al. 1999; Alcock
et al. 1999).

Estimate of $M_{\rm mIS}$ is a subtle matter. The
value depends on description of macroscopic element mixing, which is
still uncertain.
To illustrate the situation, we plot in the upper
panel of Fig.\,3 the evolutionary tracks downloaded from BASTI
library (Pietriferini et al. 2004) for metallicity parameter
$Z=0.008$, which is regarded representative for young stars in the LMC.
The tracks  were calculated in two versions,
adopting  $\alpha_{\rm ov}=0$ and 0.2 in the expression for the
overshooting distance $\lambda=H_p\alpha_{\rm ov}$, where $H_p$ is
the pressure distance scale at the core boundary. The IS boundaries
shown in Fig.\,3 refer to simultaneous instability of the first and
second overtones and they are shifted to higher temperatures by
about 0.015 in $\log T_{\rm eff}$ relative to those for the F mode.
The lower panel compares the PL relations corresponding to two
versions of the tracks at stellar masses 3 and $4M_\odot$ with the OGLE-III data
for the 1O/2O Cepheids. All pulsation properties were calculated by
us (see below) for deep envelope models at the surface parameters
taken from tracks and adopting the same input physics. Here and
in the rest of this paper, the Wesenheit index for the models is
calculated adopting 18.5 for distance modulus to LMC, which is close
to the mean value from recent determinations by various methods
(Shaefer  2008). The uncertainty of 0.05 of this value translates in
less than $0.1M_\odot$ in inferred masses of the objects and has no
bearing on our conclusions. The bolometric corrections and color indices are
taken from Kurucz (2004).

The  $M_{\rm mIS}=4.5M_\odot$ value derived from the models
calculated with $\alpha_{\rm ov}=0.2$ is similar to that determined
by Alibert et al. (1999) and Girardi et al. (2000). However, with
$\alpha_{\rm ov}=0$ the extent of the loop is not monotonic function
of mass. Instead, it reaches maximum at $M\approx4M_\odot$ and
starts increasing again at $M>5M_\odot$. Such a non-monotonic
behavior is seen also at $Z=0.01$ but not at $Z=0.004$, and we do
not know what is its origin. What matters for us here is that at $Z$
corresponding to young LMC stars,  with BASTI evolutionary tracks,
we get helium burning Cepheids of shortest periods. Still, as we
may see in the bottom panel of Fig.\,3, by far most of the 1O/2O
Cepheids in LMC have shorter periods. Are they then objects which
cross the instability strip for the first time? There are also
problems with this hypothesis. At the short periods, the first
crossing models with masses $M\lesssim3M_\odot$ and moderate
overshooting seem consistent with data. However, this is not true
for majority of the objects. Note that models of the same mass lay
on nearly the same line in the $\log P-W_I$ plane, regardless of
crossing number and the overshooting distance. We may see that most of
the stars have masses closer to 3 than $4M_\odot$ and they are
overluminous. One may consider higher $\alpha_{\rm ov}$ value but
the question is why such objects are so numerous though $\tau_{\rm
IS}$ would be over 5 times less than for the short period objects
and over 30 times less than for the $4M_\odot$ objects in helium
burning phase.

It would be easy to identify the IS crossing if we
could measure rate of period change arising from evolutionary
changes. Furthermore, such measurements would provide valuable constraints
on stellar models. Unfortunately, the fast period changes, which are
most often found in overtone Cepheids, have clearly non-evolutionary
origin (Poleski 2008).

The use of the second overtone periods allowed us to constrain parameters
of the 1O/2O Cepheids. In our study, we partially relied on the old
Warsaw codes calculating stellar models and their linear nonadiabatic oscillation
properties (see e.g. Pamyatnykh 1999).  However, these codes are not suitable
for determination of the IS boundaries because convection is treated in a crude way.
Therefore for this aim, we used a new code developed by
Smolec and Moskalik (2008a), which calculates radial mode frequencies and
growth rates for unfitted envelope models
with a more sophisticated treatment of
convection, which allows to determine the range of the instability
in a more or less reliable way. If the envelope is deep enough, the
frequencies are found very close to those calculated
using the old code for complete models with the same surface
parameters. Our evolutionary models in the first crossing phase of the IS were used
only as the reference. We varied luminosity relative to the
evolutionary models, calculated with no overshooting and looked for
the increment $\Delta L$ needed to explain periods of the 1O/2O
Cepheids. As the standard in our calculation we adopted the Asplund
et al. (2002) heavy element mixture (hereafter AGS) and the OP opacity data (Seaton 2005). Convection
is treated in the way described recently by Smolec \& Moskalik (2008a)
and as the standard, we adopt $\alpha_{\rm MLT}=1.5$. We will
briefly discuss consequences of alternative choices. There are more parameters describing
time-dependent convection. Here, we use their values adopted by Baranowski et al. (2009,
set B in Table 3) and they will not be discussed in the present paper.

\begin{figure}
\centering
\includegraphics[width=\textwidth, clip]{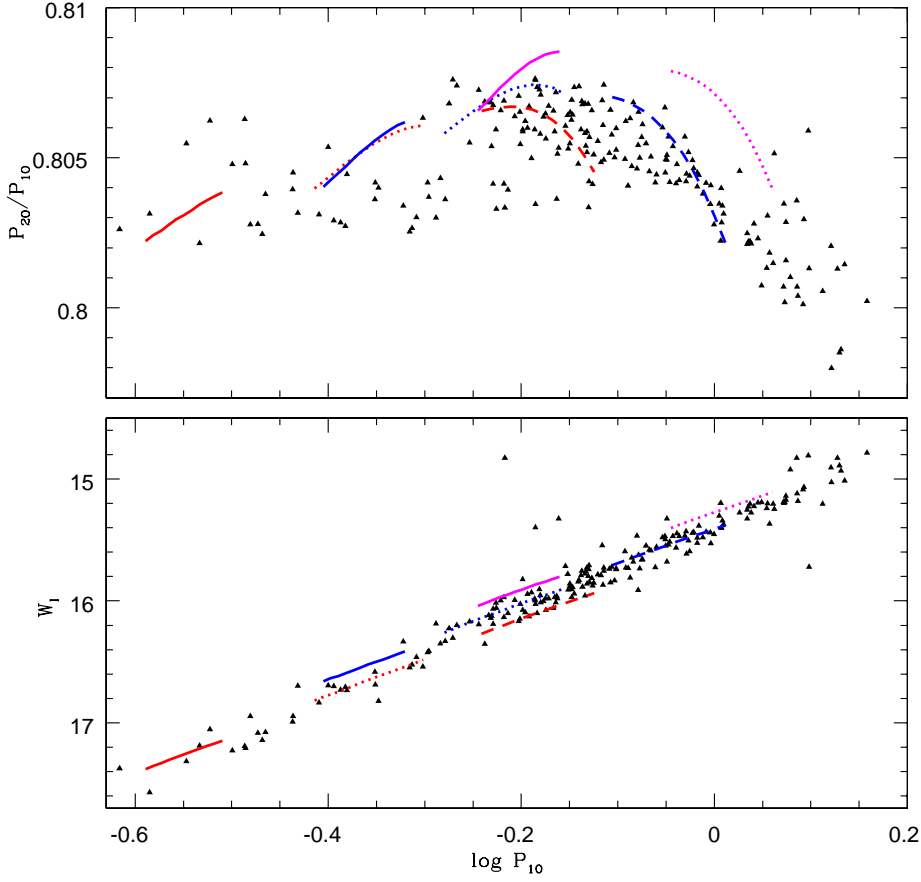}
\caption{Calculated Petersen diagram (top) and PL relations (bottom)
for selected models calculated with the metallicity parameter
$Z=0.006$ compared with the OGLE-III data for the 1O/2O Cepheids.
The segments correspond to the simultaneous instability of the 1O an
2O modes for models of specified mass and luminosity. The red, blue,
and magenta lines refer to 2.5, 3.0, and 3.5 $M_\odot$ models,
respectively. The solid lines correspond to evolutionary models
calculated  with $\alpha_{\rm ov}=0$.
Dotted and dashed lines correspond to models with artificial
luminosity increments $\Delta\log L=0.2$ and 0.4. Data are shown with
solid triangles.}
\end{figure}
The main results of our survey are shown in Figs.\,4 and 5, where calculated relations are
superimposed on the data points. Let us first focus on the PL relations shown in the bottom panels.
The effect of metallicity decrease is similar to that of luminosity increase, that is, the segments are moved
upward along the constant mass line. With only few exceptions, the points are contained between
the 2.5 and $3.5M_\odot$ lines. We may see also that beyond the short period range, a significant excess of
luminosity over the model values is required to match the data. The period ratio is indeed a sensitive
probe of the excess.
\begin{figure}
\centering
\includegraphics[width=\textwidth, clip]{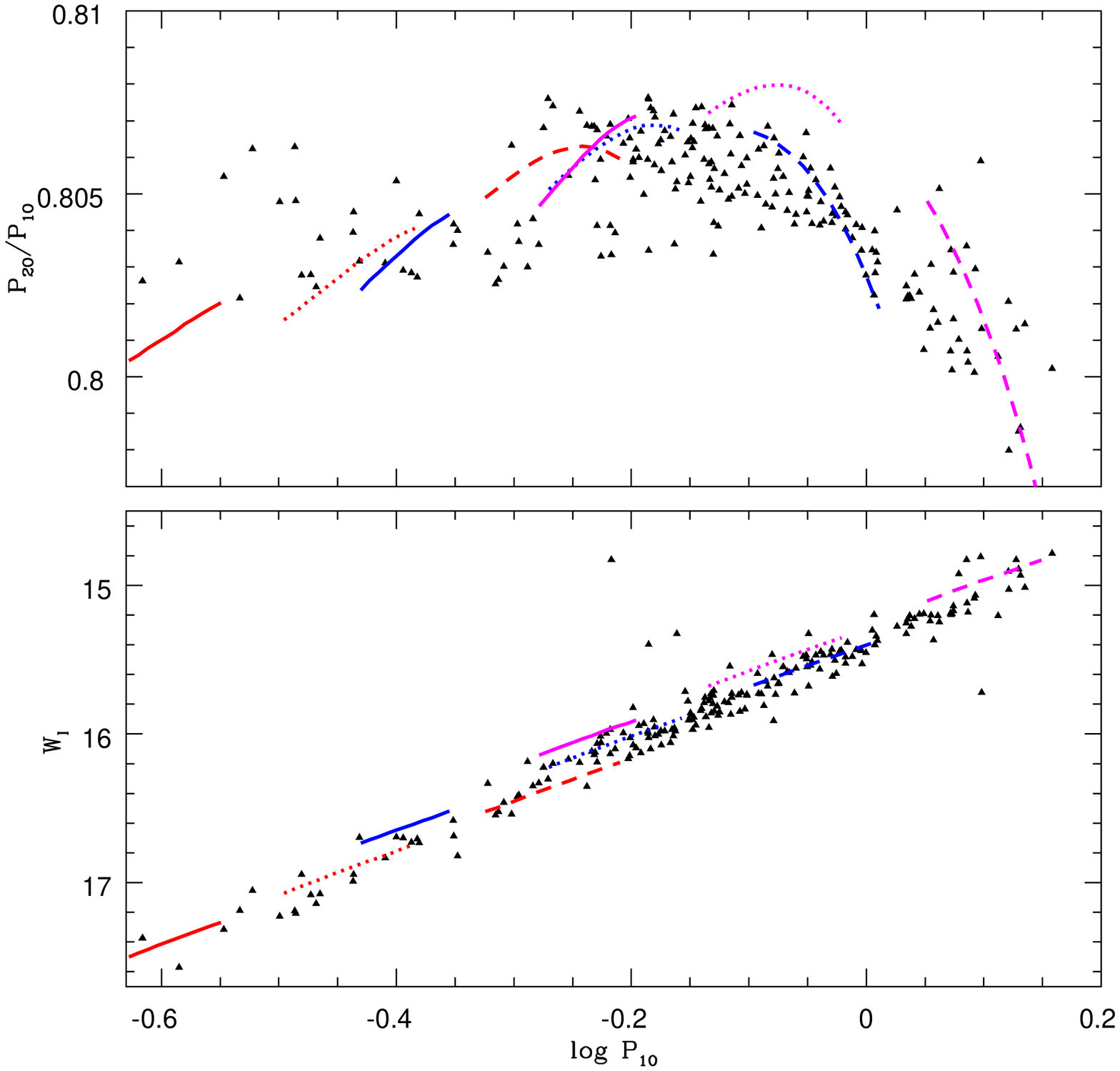}
\caption{The same as Fig. 4 but for
models calculated with the metallicity parameter $Z=0.008$.}
\end{figure}

Most of the data points in the Petersen diagram fall into the range
limited by the segments corresponding to $M=2.5$ and $3M_\odot$,
and luminosity excess $\Delta\log L=0.4$ at $Z=0.006$. Somewhat
higher values of $Z$ and/or $M$  are needed to explain data at
longest periods. Note in Fig.\,5 that at $Z=0.008$ the segment
corresponding to $M=3.5M_\odot$ and $\Delta\log L=0.4$ runs through the mid
of the data band for the longest periods. In Fig.\,4 ($Z=0.006$)
such segment is not shown because it is outside the data range. In general,
the models with $Z=0.006$ are in better agreement with data but
require larger $\Delta\log L$.  Considerable spread of points at the
shortest periods may perhaps reflect spread in the metallicity.

The $P_{1\rm O}/P_{\rm F}$ period ratios are quite sensitive to $Z$.
 Using these data, Buchler (2008) derived ranges for Cepheid
 metallicities in various galaxies.
For stars in  the LMC he found the $Z$-ranges (0.0024, 0.0093) and (0.0062,
 0.0124) with the AGS and GN93 (Grevesse and Noels 1993) mixtures, respectively. His
calculations were done with the OPAL opacity data while in our survey we
used the AGS mixture and OP data. Thus, it was important to check how use of different mixtures
and opacities affects our conclusions. We found that the $P_{2\rm O}/P_{1\rm O}$
dependence on $P_{1\rm O}$ is much less sensitive to the choice of
mixture than to opacity data. At low frequencies, where the effects
are the largest the upward shifts of the $P_{2\rm O}/P_{1\rm O}$ ratio caused by
use of the GN93 mixture is below 0.0005, while the downward shift
caused by use of the OPAL data is about 0.002, which is similar to
that caused by using $Z=0.008$ instead of 0.006. Thus, we do not see
contradiction between Buchler's and ours assessment of Cepheid
metallicity in the LMC.

The choice of $\alpha_{\rm MLT}$ has only a minor effect on
$P_{2\rm O}/P_{1\rm O}$ ratio at specified $P_{1\rm O}$. The use of $\alpha_{\rm
MLT}=1.8$ instead of 1.5 causes the 0.0005 upward shift and a
similar size downward shift is obtained with $\alpha_{\rm MLT}=1.0$.
Our conclusion regarding luminosity excess holds for different
choices of the opacity data, heavy element and the mixing length.
However, because of the significant difference of the period ratios
calculated with the two opacity data, we do not put much weight to
our assessment of the $Z$ values.

Post-main sequence models with moderate overshooting and masses between
2.5 and $3.5M_\odot$ explain data for the objects with $\log P_{1\rm O}\lesssim-0.2$.
To this category belong the seismic models of
the two 1O/2O/3O Cepheids constructed by Moskalik and Dziembowski (2005).
Certainly also the third such pulsator, which have
has similar characteristics (see Fig.\,1) may be interpreted as a standard
post-main sequence object. The same applies to the two F/1O/2O
pulsators. Note in that figure that the star at $\log P_{1\rm O}\approx0.025$ has
the period ratio significantly higher than the 1O/2O stars.
However, to explain the observed characteristics of the great majority
of the 1O/2O Cepheids in the LMC, we require a luminosity excess
larger than that caused by overshooting within normally considered range,
if indeed these objects are crossing the IS for the first time. From
our code at $M=3M_\odot$, we find $\Delta\log L=0.55\alpha_{\rm ov}$.
The corresponding increment is by some 50\% higher at
$M=3.5M_\odot$. From plots in Fig.\,3, we may find that BASTI
tracks yield somewhat higher increment resulting from overshooting
and also that the difference in $\log L$ between the first and second
crossing is about 0.25 at $M=4M_\odot$.

The large number of objects in the $M=2.5$ and $3M_\odot$ mass range
with luminosity excess $\Delta\log L=0.4$ is difficult to
explain. If they are crossing the IS before helium ignition then the
excessive overshooting may mimic rotation induced mixing.
Then, the problem to explain is why they are more numerous than the
short period objects though, with larger luminosity, they spend less time
in the IS. Alternative possibility is that the long-period 1O/2O
Cepheids represent a special population of helium burning objects
with relatively high fractional mass of the hydrogen-free core.

\begin{figure}
\centering
\includegraphics[width=\textwidth, clip]{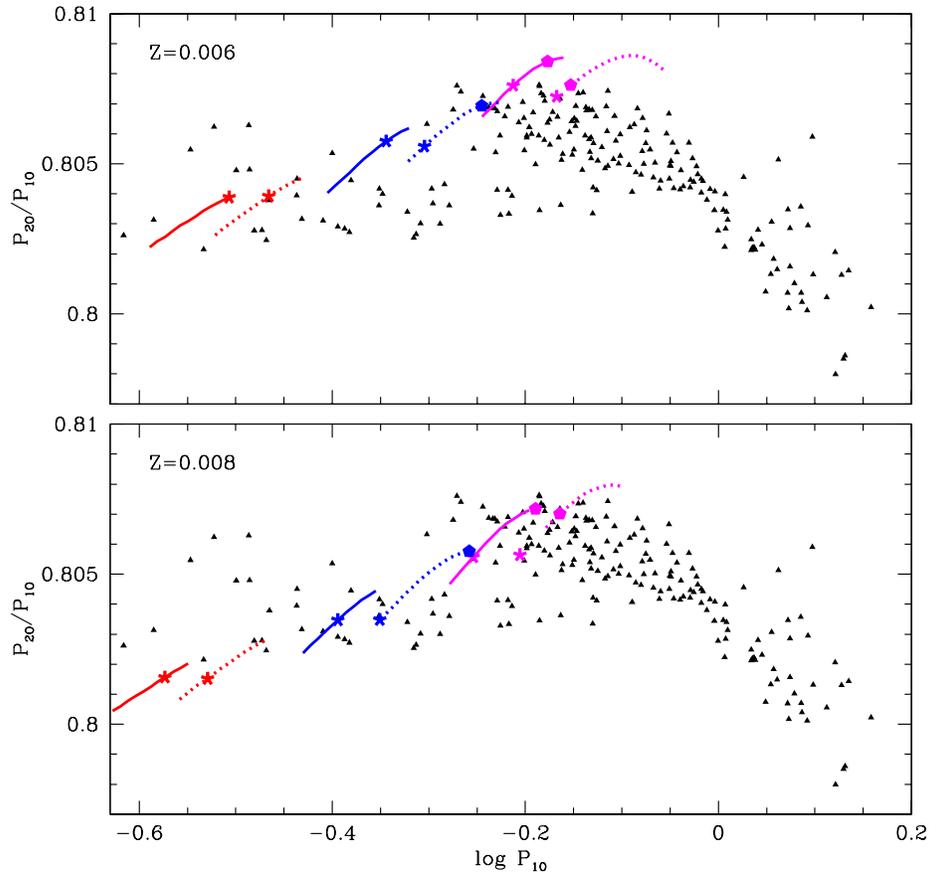}
\caption{The resonances during first crossing of the instability strip for
 stars with masses 2.5, 3.0, and $3.5 M_\odot$ (red, blue, and magenta lines, respectively).
 Solid and dashed lines refer to models calculated with $\alpha_{\rm ov}=0$ and 0.2, respectively.
 Positions corresponding to the $\nu_{5\rm O}=2\nu_{1\rm O}$ resonance are shown with the asterisk and those
 to the $2\nu_{2\rm O}=\nu_{1\rm O}+\nu_{3\rm O}$ resonance with the pentagons.}
\end{figure}

\section{Resonances and origin of double-mode pulsation}
The origin of double-mode pulsation is still lacking satisfactory
explanation (Smolec and Moskalik, 2008b). Two ways in which the
pulsation instability may lead to such a terminal state were first
discussed by Dziembowski and Kov\'acs (1984). One is a non-resonant
way acting through the feed-back effect of pulsation on mean radial
structure. The alternative way involves resonant excitation of
stable modes. A large number of papers has been devoted to the 2:1
resonance between the F and 2O modes. The coupling to stable mode
prevents the F mode to saturate driving and allows excitation of the
1O, which is linearly unstable. The effect was seen in the numerical
models but only with unrealistic parameters (Kov\'acs and Buchler
1988, Smolec 2008). The three-mode resonance $2\nu_{1\rm O}=\nu_{\rm F}+\nu_{2\rm O}$ was shown
by Smolec and Moskalik (2007) to cause double-mode pulsation in models of
$\beta$ Cephei stars. Unfortunately, we do not know any star of this type
with only two radial modes excited.

Both types of resonances, but involving higher order-overtones,
occur in certain range of our models.
In Fig.\,6, we marked positions of the resonances within (and
somewhat outside) of the IS in our models. Here, only complete
evolutionary models calculated with $\alpha_{\rm ov}=0$ and 0.2 were
used. We may see that at $\log P_{2\rm O}\lesssim-0.3$ the equality
 $\nu_{5\rm O}=2\nu_{1\rm O}$ occurs within
the IS for our models. In fact a signature of resonance centered at
$\log P_{1\rm O}\approx-0.5$ in the Fourier coefficient for the 1O
Cepheids has been reported in S2008. Inspecting plots in Fig.\,6, we
may see that this implies masses around $2.5M_\odot$ for models
calculated with $\alpha_{\rm ov}=0.2$ and somewhat higher if
$\alpha_{\rm ov}$ is less.

The resonant coupling to the strongly damped the 5O mode may cause a
significant amplitude reduction of the 1O amplitude allowing
excitation of the 2O, which is normally disfavored. This effect may
explain origin of the double-overtone pulsation in the short-period objects.
At longer periods, the resonance occurs outside of the IS.
We do not have complete models for most of the long-period objects so unfitted envelope models are
unreliable for calculation of high overtone frequencies and one could speculate that the resonance
might return to the IS in the range of the decreasing $P_{2\rm O}/P_{1\rm O}$. However, we think
that this is unlikely because we found that for the $4M_\odot$ BASTI models, which reproduce
periods at the bottom of this branch, we are far from resonance.

The $2\nu_{2\rm O}=\nu_{1\rm O}+\nu_{3\rm O}$ resonance occurs within the IS
only in a narrow intermediate period range. Thus, also this
resonance cannot explain origin of double-mode pulsation in most of
the 1O/2O Cepheids. Such form of pulsation must predominantly
arise from non-resonant saturation of the driving effect. Only
nonlinear modeling may tell us how does it happen. The histogram in
Fig.\,2 yields an important hint for such a modeling. Selecting a
model with the first overtone period near 0.7 d, we are very likely
to see the instability developing to double-overtone pulsation.

The explanation in terms a non-resonant saturation seems difficult for
the two 1O/3O Cepheids (Soszy\'nski et al. 2008). In these cases,
the 2:1 resonance between the second and the seventh overtone may play a role.
The resonant coupling to this strongly damped mode may totally prevent excitation
of the second overtone.

\section{Conclusion and discussion}

We confronted the OGLE-III data on 203 LMC Cepheids, in which
first and second overtones are excited (the 1O/2O Cepheids), with
model calculations. The data used in our paper include periods and
the Weseinheit index, $W_I$, which combined with the LMC distance is
used as the measure of the absolute magnitude.

Unlike the F/1O, the 1O/2O Cepheids cannot be regarded an
exceptional form of stellar pulsation. Not only the latter form
occurs three times more often among the LMC objects  but,  more importantly, in a
certain period range around 0.7d, it is the most common form. The
data on the 1O/2O Cepheids present interesting challenges to stellar
evolution and pulsation theory.

The period-luminosity ($W_I - \log P_{1\rm O})$ relation
constrains masses of the objects. Nearly all of them fall in the
2.5 to $3.5M_\odot$ mass range. The Petersen diagram ($P_{2\rm O}/P_{2\rm O} - \log P_{1\rm O}$)
yields constrains on metallicity and luminosity. We focused on the former, adopting for the metallicity
parameters $Z=0.006$ and 0.008. The inferred luminosities for the
objects with the first overtone period shorter than about 0.6 d are
well explained with post-main sequence stellar models crossing the
instability strip for the first time, calculated assuming no or
moderate overshooting.

However, for objects with longer periods,
which constitute majority of the sample, we find significant
luminosity excess. If these are objects crossing the instability strip before
helium ignition then the excess could be due to an overlarge
overshooting, perhaps connected with fast rotation. The difficulty
of this explanation is the short crossing time
of the instability strip while the number of the objects showing the
luminosity excess is relatively high.
If these are helium burning objects then the difficulty is the low inferred mass.
Standard evolutionary track for stars
with $M\lesssim3.5$ and acceptable metallicity do not enter the
instability strip in this evolutionary phase. We must postulate a
non-standard evolution leading to objects with relatively massive helium cores.
This could be a large mass loss in the red giant phase or again a large overshooting in the
main sequence phase. The former option would be plausible if the stars were in
binary systems for which we do not have yet any evidence. The latter option remains
to be verified by means of new stellar evolutionary calculations. The result is not
clear. As a warning, we have to recall that in certain mass range, BASTI
models calculated without overshooting entered the instability strip
in helium burning phase while those with  overshooting did not (see Fig.\,3).

The challenge to stellar pulsation theory is to identify conditions
leading to excitation of two rather than one pulsation mode and
to explain why the 1O/2O pulsators are more frequent than F/1O and also
2O. We found two types of resonances, which may promote double-overtone
pulsation but only in the case of the short-period objects. For the majority of
stars, such pulsation may result only from non-resonant
saturation. In a certain period range this must be the preferred form of
terminal development of the instability. The question why it is so may be answered only
with the help of nonlinear modeling of stellar pulsation.

\Acknow{We thank Igor Soszy\'nski and Rados{\l}aw
Poleski for providing us data used in our plots. We also thank Igor
Soszy\'nski and Alosha Pamyatnykh for reading preliminary version of
this paper and suggesting improvements. This work has been supported
by the Polish MNiI grant No 1 P03D 011 30.}


\begin{references}
\refitem{Alcock, C.,Allsman, R.A., and Alves, D.R.}{1999}{\ApJ}{117}{920}

\refitem{Alibert, Y., Baraffe, I., Hauschildt, P., and Allard, F.} {1999}
{\AA}{344}{551}

\refitem{Asplund. M., Grevesse N., Sauval A.J., Allende Pieto, C., and
Kisleman D.} {2004} {\AA} {417} {751}

\item{Baranowski, R., Smolec, R., Dimitrov, W., et al.}{2009}{~}{~}{submitted to \MNRAS}

\refitem{Buchler, J. R.}{2008}{\ApJ} {680} {1412}

\refitem{Buchler, J. R. and Szab\'o, R.}{2007}{\ApJ} {660} {723}

\refitem{Dziembowski, W.A. and Kov\'acs, G}{1984}{\MNRAS}{196}{731}

\refitem{Girardi, L., Bressan, A., Bertelli, G., and Chiosi, C.}
{2000} {\AAS}{141}{371}

\item{Grevesse, N. and Noels, A., 1993 in  Origin and Evolution of the Elements,
    eds. Pratzo M., Vangioni-Flam E., and Casse M.,
    Cambridge Univ. Press, p. 15}

\refitem{Iglesias, C.A. and Rogers, F.J.} {1996} {\ApJ} {464} {943}

\refitem{Koll\'ath, Z., Beaulieu, J.P., Buchler, J.R., and Yecko P.}
{1998} {\ApJ} {502} {L55}

\refitem{Kov\'acs, G. and  Buchler, J. R.}{1988}{\ApJ}{308}{661}

\item{Marquette, J.M., Beaulieu, J.P., Buchler, J.R., et al.} {2009}
{ArXiv:0901.0995v1 [astro-ph]}

\refitem{Moskalik, P. and Dziembowski, W.A.} {2005}{\AA}{434}{1077}

\item{Moskalik, P., Ko{\l}aczkowski, Z., and Mizerski, T.} {2004}
{in Variable Stars in the Local Group, ed. D.W. Kurtz, and K.
Pollard, ASP Conf. Ser., Vol. 310, 498}

\item{Kurucz R. L.}{2004} {http://kurucz.harvard.edu}

\refitem{Pamyatnykh, A.A.} {1999} {\Acta} {49} {119}

\refitem{Petersen, J.O.} {1973} {\AA} {27} {89}

\refitem{Pietrinferni, A., Cassisi, S., Salaris, M., and Castelli, F.} {2006} {\ApJ}
{642}{797}

\refitem{Poleski, R.}{2009}{\Acta}{58}{313}

\refitem{Schaefer, B.E.}{2008}{\AJ}{135}{112}

\refitem{Seaton, M.}{2005}{\MNRAS}{362}{L1}

\refitem{Simon, N.R.}{1982}{\ApJ}{260}{L87}

\item{Smolec, R.} {2008} {ArXiv:0803.1441 [astro-ph]}

\refitem{Smolec, R. and Moskalik, P.}{2007}{\MNRAS}{377}{645}

\refitem{Smolec, R. and Moskalik, P.}{2008a}{\Acta}{58}{193}

\refitem{Smolec, R. and Moskalik, P.}{2008b}{\Acta }{58}{233}

\refitem{Soszy\'nski, I., Poleski, R., Udalski, A., Kubiak, M., Szyma\'nski, M.,
Pietrzy\'nski, I., Wyrzykowski, L., Szewczyk, O., and Ulaczyk, K. }{2008a}{\Acta}{58}{163}

\refitem{Soszy\'nski, I., Poleski, R., Udalski, A., Kubiak, M., Szyma\'nski, M.,
Pietrzy\'nski, I., Wyrzykowski, L., Szewczyk, O., and Ulaczyk, K. }{2008b}{\Acta}{58}{163}

\refitem{Szab\'o, R., Koll\'ath, Z. and Buchler, J.R.} {2004}{\AA} {425} {627}
\end{references}
\end{document}